\documentstyle[12pt,openbib,times]{article}
\newcommand{\f}{\frac}

\def\be{\begin{equation}}
\def\ee{\end{equation}}

\begin{document}
\date{}
\title{High Density Strange Star Matter and Observed Parity
Doubling of Excited Hadrons}
\author{Taparati Gangopadhyay $^{1,*}$, Manjari Bagchi
$^{1}$, \\Mira Dey $^{1,\dagger}$, Jishnu Dey $^{1,\ddagger}$ and
Subharthi Ray ${^2}$}

\maketitle

\noindent $^1$ Dept. of Physics, Presidency
College, 86/1 , College Street, Kolkata 700 073, India. \\
$^2$ Inter University Centre for Astron. \& Astrophys., Post Bag
4, Ganeshkhind, Pune 411 007, India.
\\$^*$ Ramanna Junior Research Fellow, DST, Govt. of
India\\$^{**} $ Ramanna Research Associate, DST, Govt. of India\\
$\dagger$ Ramanna
Emeritus Fellow, DST, Govt. of India, Associate, IUCAA\\
$\ddagger$ CSIR Emeritus Scientist, Govt. of India; Associate,
IUCAA.

\begin{abstract}{Parity doubling is observed in hadron states
around 1.5 GeV - and the estimated energy density is found to be
high. When a large excitation energy is available and the pions
decouple from the quarks, the QCD interaction is still not
perturbative. And signature of such a system can exist in the
form of small ratio of the shear viscosity to entropy density.
This is true for an equation of state which is applicable to the
surface of strange star. We indicate the correspondence of parity
doubling with the apparently disconnected model calculation of
compactness of some pulsars. }\end{abstract} \vskip .5cm

\vskip .5cm

\noindent Keywords(pacs) {11.30.Rd} {Chiral symmetries} {11.10.St}
{Bound and unstable states} {11.15.Tk} {Other nonperturbative
techniques}

\section{Introduction}

Taking the recently determined experimental hadronic resonances
we see that parity doubling occurs at energy densities which are
comparable to densities of strange stars. Parity doubling is an
indicative of chiral symmetry restoration (CSR), which can be
also be found in strange star models, that has successfully
explained many inexplicable features of the astrophysical compact
objects.

In this model, which we call ReSS, the u, d and s quarks become
gradually less massive as the density increases~\cite{D98, AandA}.
Pions decouple from quarks and CSR is observed partially in the
surface and strongly at the core of the star \cite{njl}. Since
existence of strange stars cannot be directly proved with the
present day observation technique, the support of CSR from parity
doubling in hadronic resonances may play an important role. In an
earlier work by Bagchi et al. \cite{njl} the density dependent
mass of ReSS was included in their paper following the well known
model due to Nambu and Jona Lasinio. There the pion coupling to
the quarks namely $f_\pi$ was found to decrease signaling
decoupling of the pion from the quarks.

The parity doubling in baryon resonances was rediscovered in 2000
by Glozman \cite{gloz} and there are many follow up papers
\cite{gloz2} - including \cite{jaffe}. In the last mentioned
paper, the authors think that chiral symmetry realized in the
Nambu Goldstone mode does not predict the existence of degenerate
multiplets of hadrons of opposite parity. However they assert
that their arguments do not preclude the restoration of chiral
symmetry at high temperature or high energy density. In the
present paper we seek to show that indeed the energy densities
involved are comparable to ReSS and are thus expected to lead to
CSR. To do this we show the correspondence to a situation where
the chemical potential is high in a strange quark matter (SQM)
equation of state (EoS). This EoS, along with a density dependent
quark mass, has been used to construct strange star models at
zero~\cite{D98} and finite temperature~\cite{AandA}.

This strange star model, in contrast to the recent papers seeking
to establish parity doubling, consists of a large $N_c$ mean field
approximation with a realistic modified Richardson potential
between quarks, but is crucial in explaining many observations
like (a) super bursts~\cite{monika}, (b) the existence of minimum
magnetic field for all observed pulsars~\cite{raka}, and (c)
absorption and emission bands along with high
redshift~\cite{manjari1}.

We observe that the list of states from 2 GeV to 2.51 GeV for the
mesons, which show the so called clustering observed recently by
Afonin~\cite{af}, have very high energy density. This suggests
CSR, which is built-in for our model. With respect to the
objections to CSR~\cite{jaffe}, we agree with Afonin who observes
``one should be careful with any statements forbidding CSR which
are based on the language of the low - energy field theory like in
... ". Most of the resonances are recently found and need
confirmation according to the Particle Data Group (PDG) and we
emphasize that confirmation and determination of new states is
important since this is related to CSR.

The possibility that parity doubling is observed in baryon
resonances was realized by Dey and Dey in 1993~\cite{dd}. This
was inspired by the early work of Barut~\cite{bar1} where he had
looked at parity doubled states in the conformal O(4,2) model as
early as 1965.

There have been many models for indicating why parity doublets
appear in QCD. The most recent one is the toy model of Cohen and
Glozman~\cite{hep-ph0512185} in which there are infinite number of
pions and $\sigma$-s and which is claimed to mimic large $N_c$
QCD. Before this, there was the question of parity doubling in the
baryons treated by Jaffe, Pirjol and Scardicchio~\cite{jaffe}. And
finally there was the revival of the model of Ademollo, Veneziano
and Weinberg \cite{avw} by Afonin \cite{af}.

In section~\ref{reso} we deal with the baryonic resonances at
high energy. In section~\ref{sseos} we refer to strange stars and
low viscosity-entropy ratio which has relevance to the recent
relativistic Heavy ion collision (RHIC) experiments. We summarize
and conclude in section~\ref{sumcon}.

\section{Resonances at high energy}\label{reso}

The large $N_c$ ReSS model employs a potential which has
asymptotic freedom with a scale parameter $\Lambda = 100$ MeV and
confinement scale of $\Lambda^\prime = 350$. The stars are fitted
with a density dependent form for quark masses at zero as well as
finite temperature~\cite{D98,AandA} after solving for $\beta$
equilibrium and charge neutrality for u, d and s quarks and
electrons. The hydrostatic equilibrium equation (TOV equation) is
then solved self consistently to find the properties of the
strange stars at all radii, the density varies from about 4.5 to
15 times normal nuclear density for a maximum mass of about 1.5
$M_\odot$.

We must point out that this  SQM model fits the ground state
baryons and their magnetic moments~\cite{manmag1, manmag2}
without the pion degrees of freedom. Also the quarks are found to
decouple from the pion~\cite{njl} as already stated.

Coming to the experimental states, we note that there are 314 even
parity mesons and 308 for odd parity in the range $\Delta ~E$ = 2
to 2.51 GeV \cite{pdg}. Specific examples of parity doubling are
striking~: for example, $b_1 ~(I^G(J^{PC}) 1^+(1^{+ -})1960$,
$\rho ~1^+(1^{- -})1965$. Interestingly, mesonic spectrum can be
predicted for example from the discrete quark - gluon plasma
states of a finite bag as shown by Dey, Tomio and Dey \cite{dtd}.
With a reasonable bag pressure B given by $B^{\f{1}{4}}~=~200~MeV$
one can estimate the radii of these mesons in the bag model. We
also note that in the 2006 PDG tables there are some states
marked X which could soon be cleared up. What is striking is that
even and odd parity states come up together. The way the density
of states is calculated is standard \cite{dtd}, the entropy is
minimized and the Laplace transform of its second derivative
leads to the density of states in saddle point approximation. We
refer the interested reader to the original paper. The density of
states lead to a limiting temperature which is given in a Table in
Dey, Tomio and Dey~\cite{dtd} to be 143.7 for a meson mass 2.1
$GeV$ and 142.8 for 2.5 $GeV$ and the radii are calculated in the
same Table to be 0.844 $fm$ and 0.895 respectively. We will see
that these match with the hadron radii calculated from other
simple considerations. Such massive resonances in such small
radii lead naturally to very high matter density.

Most of these states are new and we tabulate in
Table~(\ref{list}). The excited states of the pions are wrongly
marked as iso-singlets in the original Tables~\cite{pdg} but Dr.
Eidelman assures us that these will be corrected in
the 2007 edition of PDG.

We now estimate the radii of high energy mesons. Let the energy
and average radius respectively be $E$, $R_{reson}$ for a mesonic
resonance. Then from the surface energy density of a ReSS of
$\epsilon~=~627.36 ~MeV/ fm^{3}$ one gets

\begin{equation}
\f{E}{\epsilon} = {\f{4 \pi }{3}}R_{reson}^3.
\end{equation}
$R_{reson}$  is 0.913 to 0.977 $fm$ when E is 2 $GeV$ or 2.5
$GeV$ respectively.

The calculations that we present for the strange star are very
straightforward. At the surface of the star, where the number
density $n_q$ is minimum (about 4.6 times the normal nuclear
density), on the average the quarks occupy a sphere of radius
$r_n~=~0.51$ $fm$ or less. We notice that the surface area of the
star at any radius $r$ is $4 \pi r^2$ whereas the projection of a
quark with an effective volume $V = \f{4}{3} \pi r_n^3$ is just
$\pi (2~r_n)^2$. This gives us the inequality
$r_n^2~\le~r^2/n_q$. Since we know the number density $n(r)$ at
any $r$ we can get $n_q~=~4~\pi~r^2~2 r_n n(r)$ for a thin shell
of depth $ 2 r_n$. We get the above number when the inequality is
assumed to be saturated and we put the surface number density
appropriate for a star of mass  $\sim 1.5~M_\odot$. The smallness
of $r_n$ justifies the thin shell approximation.

Inside the star where the density is about 15 times the normal
nuclear density, the number reduces to 0.314 $fm$ showing that
the effective radius of the quarks gets closer to a partonic
picture since the chiral symmetry is restored in the model with a
density dependent quark mass~\cite{D98}.

The mean inter-particle distance $r_0~=~0.47~ fm$ at the surface
of a strange star, assuming quarks have an effective volume
$\f{4}{3}\pi~r_0^3$ and the usual colour, 3-flavour and spin
degeneracy. And like $r_n$, at the centre of the star $r_0$
decreases to $0.315~fm$.

Thus we see that the average radius of the resonance is close to
twice the size of $r_n$ or $r_0$. Indeed the energy density is
slightly larger than that at the surface. It also corresponds to a
very little depth inside the star from the surface.  These radii
agree with the estimates of Dey, Tomio and Dey~\cite{dtd} which
uses bag model with finite size corrections. But the number of
states given by the bag model, which is 6674 between 1700 and 2100
$MeV$ is too numerous~\cite{ddt} and may be due some bag
artifact - or due to many resonances being so far unobserved.

We do a completely different calculation to demonstrate that for
baryons also one is dealing with high density system. All models
employing group theory seem to give a spectra which looks like a
rotational band~\cite{toki, dd}. We can see that the parity
doubling occurs for isobars with a 3/2+ state at $\sim 1705 ~MeV$
(observed by Manley and Saleski~\cite{man} and Li et al.~\cite{li}
and a 3/2- state at 1700 $MeV$ \cite{pdg}. We take the 1700  3/2+
state as a band head and look for rotational states above it. The
L = 2 states are well known as the quartet states 7/2+ (1950),
5/2+ (1905), 3/2+ (1920) and 1/2+ (1910). For L = 4 there are only
three states : 11/2+ (2420), 9/2+ (2300) and 7/2+ (2300), the
fourth 5/2 + state is worth looking for. Apart from the L=6,
15/2+ state there could be a 13/2+ state also at 3230 $MeV$. The
even parity isobar $15/2^+$ is at 2950 $MeV$.

Using a simple moment of inertia model with the $\alpha~=~\hbar^2
/2I$ with $I~=~\f{2}{5} M R_{reson}^2$ one gets the resonance
radius $R_{reson}~=~ 0.96~fm$ and the moment of inertia is
$I~=~627~ MeV~fm^2$. The quartet state comes out at 1886, the L=4
at 2321 and L=6 states at 3004. For this L=6 state the energy
density comes out to be $\epsilon~=~ 810~MeV/fm^3$ comparable to
an energy density observed well inside a strange star. This is of
some interest since the strange star surface marks the onset of a
different phase. Taking the $\Lambda(2350)$ 9/2+ state with L=4 we
get $I~=~518~MeVfm^2$, $R_{reson}~=~0.9~fm$ and energy density
$\epsilon~=~770 MeV/fm^3$.

The resonances have widths so we deal with the centroids and
attempts to more accurate fitting would be futile. On the other
hand this approximate but simple calculation seems to work.

\section{Strange matter, strange stars and low viscosity-entropy ratio}\label{sseos}

With the value of $r_0$ at the surface for a ReSS at T = 80
$MeV$, we satisfy the remarkable inequality $4~\pi~\eta/s~\ge~ 1$
\cite{lgbddsr}. This was invoked in a much quoted paper by
Kovtun, Son and Starinets \cite{kss} (KSS). Here $s$ stands for
the entropy density.

The relevant arguments of KSS~\cite{kss} are very appealing to us
since it only invokes general principles like Heisenberg
uncertainty relation for the typical mean free time of a
quasi-particle and $s$ which in turn is proportional to the
density of the quasi-particles. From here to our model is just one
short step of identifying the quasi-particles to be the dressed
quarks of the mean field description for a large colour effective
theory.

The matter in the strange star is a  so called perfect interacting
liquid where bound reaches the fraction ${(4\pi)}^{-1}$ and thus
it may be the same fluid which Lacey marks as RHIC in  figure 3 of
his paper - which stands for relativistic heavy ion
collisions~\cite{lacey}. The point is that in RHIC one obtains a
large elliptic flow which demands a very small shear viscosity
whereas perturbative QCD yields a value which is almost ten times
the bound.

We thus see a possible connection between parity doubling and high
energy matter at the surface of a strange star. Then the fact
that the strange matter supports such a low shear viscosity
constraint relates it to RHIC. Considerable understanding of QCD
will result from more definite observational signatures for the
existence of strange stars, more experimental study of parity
doubling in hadrons and explanation of RHIC data.

\begin{table}
\centering \caption {List of mesons at high energy which show that
including the degeneracy (2J+1)(2T+1) the total number of odd and
even parity states match roughly in the range 2 to 2.51 $GeV$.}
\begin{tabular}{|c|c|c|c|}
\hline
$I^G(J^{PC})$ & $State$ & $I^G(J^{PC})$ & $State       $\\
\hline
       $0^+(2^{++})$ & $f_2(2000) $ & $1^+(1^{--})$ & $\rho (2000) $\\
       $0^+(0^{++})$ & $f_0(2020) $ & $1^-(2^{-+})$ & $\pi_2 (2005)$\\
       $1^-(0^{++})$ & $a_0 (2020)$ & $0^+(0^{-+})$ & $\eta (2010) $\\
       $1^+(3^{+-})$ & $b_3(2025) $ & $1^-(1^{-+})$ & $\pi_1(2015)$ \\
       $1^-(4^{++})$ & $a_4 (2040)$ & $0^-(3^{--})$ & $h_3(2025)$ \\
       $0^+(3^{++})$ & $f_3 (2050)$ & $0^+(2^{-+})$ & $\eta_2 (2030)$ \\
       $0^+(4^{++})$ & $f_4 (2050)$ & $1^-(0^{-+})$ & $\pi (2070)$ \\
       $0^+(0^{++})$ & $f_0 (2060)$ &&   \\
       $1^-(3^{++})$ & $a_3(2070) $ &&   \\
       $1^-(2^{++})$ & $a_2 (2080)$ &&   \\
       $1^-(1^{++})$ & $a_1 (2095)$ &&   \\
\hline
      $0^+(0^{++})$ & $f_0(2100)$ & $0^+(0^{-+})$ & $\eta (2100)$\\
      $0^+(2^{++})$ & $f_2(2150)$ & $1^-(2^{-+})$ & $\pi_2(2100)$\\
      $0^-(2^{++})$ & $a_2(2175)$ & $0^-(1^{--})$ & $\omega(2145)$\\
      &                           & $1^+(1^{--})$ & $\rho(2150) $ \\
      &                           & $0^+(0^{-+})$ & $\eta(2190)$ \\
      &                           & $0^-(2^{--})$ & $\omega_2(2195)$ \\
\hline
      $0^+(0^{+-})$ & $f_0(2200)$ & $1^+(2^{--})$ & $\rho_2(2240)$\\
      $0^-(1^{+-})$ & $h_1(2215)$ & $0^-(1^{--})$ & $\omega (2205)$ \\
      $1^+(1^{+-})$ & $b_1(2240)$ & $0^-(0^{-+})$ & $\eta(2225)$  \\
       $1^-(1^{++})$ & $a_1(2270)$& $1^+(4^{--})$ & $\rho_4(2240)$ \\
       $1^-(2^{++})$ & $a_2(2270)$& $1^-(2^{-+})$ & $\pi_2(2245)$ \\
       $0^-(3^{+-})$ & $h_3(2275)$& $0^+(2^{-+})$ & $\eta_2(2250)$ \\
       $1^-(4^{++})$ & $a_4(2280)$ & $1^-(4^{-+})$ & $\pi_4(2250)$ \\
      $0^+(2^{++})$ & $f_2(2300)$& $1^+(3^{--})$ & $\rho_3(2250)$ \\
       $0^+(3^{++})$ & $f_3(2300)$& $0^-(4^{--})$ & $\omega_4(2250)$  \\
       $0^-(4^{++})$ & $ f_4(2300)$& $0^-(3^{--})$ & $\omega_3(2255)$ \\
      &             & $1^+(1^{--})$ & $\rho(2265)$ \\
      &             & $0^+(0^{-+})$ & $\eta(2280)$\\
      &             & $1^+(1^{--})$ & $\rho(2280)$ \\
      &             & $0^-(3^{--})$ & $\omega_3(2285)$\\
      &              & $1^+(3^{--})$ & $\rho_3(2300)$ \\
\hline
      $0^+(1^{++})$ & $ f_1(2310) $&$0^+(4^{-+})$ & $\eta_4(2320)$  \\
      $1^-(3^{++})$ & $ a_3(2310) $&$ 0^+(1^{--})$ & $\omega(2330)$ \\
      $0^+(0^{++})$ & $ f_0(2330) $&$ 1^+(5^{--})$ & $\rho_5(2350)$\\
      $1^-(1^{++})$ & $ a_1(2340) $&$ 1^-(0^{-+})$ & $\pi(2360)$\\
      $0^-(2^{++})$ & $ f_2(2340) $&&\\
      $1^-(6^{++})$ & $ a_6(2450) $&&                            \\
      $0^+(6^{++})$ & $ f_6(2510) $&&                            \\
\hline
\end{tabular}
\label{list}
\end{table}

\section{Summary and conclusion}\label{sumcon}

Recent  developments have made the rather perplexing connection
to the idea of a limit for low shear viscosity $\eta$ to entropy
density $s$, $\eta/s < 1/(4 \pi)$ to RHIC phenomenology
\cite{kss}. The limit is not satisfied for perturbative QCD
\cite{ArnMooreYaffe} and the viscosity is some ten times larger.
For matter in the centre of a strange star the same conditions
apply and the results are consistent. But at the surface of a
strange star the limit is satisfied at T = 80 $MeV$, where the
free energy has a minimum and the pressure is zero critically
ensuring a self bound star~\cite{lgbddsr}.

To summarize, parity doubling in excited state spectrum of the
hadrons is a high density phenomenon. It matches with ReSS model
indicating that quarks decouple from pions. One finds a
similarity in simple high density matter calculations rather than
in complicated low energy models. The model supports the Kovtun,
Son and Starinets bound for the ratio of shear viscosity to
entropy density at a temperature of 80 $MeV$ at the surface of a
strange star~\cite{lgbddsr}. The elliptic flow observed at RHIC is
believed to be due to such low shear viscosity. And one does not
get such flow from partonic matter which uses perturbative QCD
but rather from interacting QCD which produces dressed
quasi-particle like objects.

We conclude suggesting that the high energy density  hadronic
resonances should be explored more extensively since they support
the chiral symmetry restoration in models of strange stars and in
RHIC.

\section*{Acknowledgments}

TG, MB, MD and JD acknowledge the hospitality during visits at
IUCAA, Pune and HRI, Allahabad, India, and two of us (JD and MD)
also acknowledge a fruitful visit to ECT, Trento, Italy in 2006.
The authors are grateful to F. Iachello for some important
suggestions regarding this work.




\end{document}